\documentclass[default,iicol]{sn-jnl}

\usepackage{graphicx}%
\usepackage{caption}
\usepackage{subcaption}
\usepackage{multirow}%
\usepackage{amsmath,amssymb,amsfonts}%
\usepackage{amsthm}%
\usepackage{mathrsfs}%
\usepackage[title]{appendix}%
\usepackage{xcolor}%
\usepackage{textcomp}%
\usepackage{manyfoot}%
\usepackage{booktabs}%
\usepackage{algorithm}%
\usepackage{algorithmicx}%
\usepackage{algpseudocode}%
\usepackage{listings}%
\usepackage{placeins}

\usepackage{pgfplots} 
\usepackage{pgfgantt}
\usepackage{pdflscape}
\pgfplotsset{compat=newest} 
\pgfplotsset{plot coordinates/math parser=false}
\usepackage{hhline}
\usepackage{color}
\usepackage{soul}
\usepackage{nomencl}
\makenomenclature
\usepackage{environ}
\usepackage{tikz}
\usetikzlibrary{trees,calc,shadings,shapes.gates.logic.US,positioning,arrows}
\makeatletter
\newsavebox{\measure@tikzpicture}
\NewEnviron{scaletikzpicturetowidth}[1]{%
  \def\tikz@width{#1}%
  \def\tikzscale{1}\begin{lrbox}{\measure@tikzpicture}%
  \BODY
  \end{lrbox}%
  \pgfmathparse{#1/\wd\measure@tikzpicture}%
  \edef\tikzscale{\pgfmathresult}%
  \BODY
}
\makeatother

\usepackage{makecell}

\newcommand{\eq}[1]{Eq.~(\ref{#1})}
\newcommand{\figref}[1]{Fig.~\ref{#1}}

\begin{document}
\title[Vertiport Navigation Requirements and Multisensor Architecture Considerations for Urban Air Mobility]{Vertiport Navigation Requirements and Multisensor Architecture Considerations for Urban Air Mobility}
\author*[]{\fnm{Omar} \sur{Garc\'ia Crespillo}}
\author*[]{\fnm{Chen} \sur{Zhu}}
\author[]{\fnm{Maximilian} \sur{Simonetti}}
\author[]{\fnm{Daniel} \sur{Gerbeth}}
\author[]{\fnm{Young-Hee} \sur{Lee}}
\author[]{\fnm{Wenhan} \sur{Hao}}
\affil[]{\orgdiv{Institute of Communication and Navigation}, \orgname{German Aerospace Center (DLR)}, \orgaddress{\street{Muenchenerstr. 20}, \city{Oberpfaffenhofen}, \postcode{82234}, \country{Germany}}}
\abstract{Communication, Navigation and Surveillance (CNS) technologies are key enablers for future safe operation of drones in urban environments. However, the design of navigation technologies for these new applications is more challenging compared to e.g., civil aviation. On the one hand, the use cases and operations in urban environments are expected to have stringent requirements in terms of accuracy, integrity, continuity and availability. On the other hand, airborne sensors may not be based on high-quality equipment as in civil aviation and solutions need to rely on tighter multisensor solutions, whose safety is difficult to assess.
In this work, we first provide some initial navigation requirements related to precision approach operations based on recently proposed vertiport designs. Then, we provide an overview of a possible multisensor navigation architecture solution able to support these types of operations and we comment on the challenges of each of the subsystems. Finally, initial proof of concept for some navigation sensor subsystems is presented based on flight trials performed during the German Aerospace Center (DLR) project HorizonUAM.}
\maketitle
\nomenclature{ABAS}{Airborne-based Augmentation System}
\nomenclature{ARAIM}{Advanced Receiver Autonomous Integrity Monitoring}
\nomenclature{ATM}{Air Traffic Management}
\nomenclature{CNS}{Communication, Navigation and Surveillance}
\nomenclature{DLR}{German Aerospace Center}
\nomenclature{DTED}{Digital Terrain Elevation Data}
\nomenclature{EASA}{European Union Aviation Safety Agency}
\nomenclature{EUROCAE}{European Organization for Civil Aviation Equipment}
\nomenclature{FATO}{Final Approach and Take-off Area}
\nomenclature{FTE}{Flight Technical Error}
\nomenclature{GAST}{GBAS Approach Service Type}
\nomenclature{GBAS}{Ground-based Augmentation System}
\nomenclature{GNSS}{Global Navigation Satellite System}
\nomenclature{GPS}{Global Positioning System}
\nomenclature{HAL}{Horizontal Alert Limit}
\nomenclature{HPE}{Horizontal Position Error}
\nomenclature{ILS}{Instrument Landing System}
\nomenclature{IMU}{Inertial Measurement Unit}
\nomenclature{INS}{Inertial Navigation System}
\nomenclature{ISA}{International Standard Atmosphere}
\nomenclature{JARUS}{Joint Authorities for
Rulemaking on Unmanned Systems}
\nomenclature{MSL}{Mean Sea Level}
\nomenclature{NAA}{National Aviation Authority} 
\nomenclature{NSE}{Navigation System Error}
\nomenclature{OSO}{Operational Safety Objectives}
\nomenclature{PDE}{Path Definition Error}
\nomenclature{QFE}{Atmospheric pressure at aerodrome 
elevation (or at runway threshold)}
\nomenclature{QNE}{Reading on an altimeter on landing when the standard pressure setting is set}
\nomenclature{QNH}{Altimeter sub-scale setting to obtain 
elevation when on the ground}
\nomenclature{RPAS}{Remotely Piloted Aircraft Systems}
\nomenclature{RTCA}{Radio Technical Commission for Aeronautics}
\nomenclature{SA}{Safety Area}
\nomenclature{RTODV}{Rejected Take-off Distance for VTOL-capable aircraft}
\nomenclature{SAIL}{Safety Assurance and Integrity Levels}
\nomenclature{SBAS}{Satellite Based Augmentation System}
\nomenclature{SORA}{Specific Operations Risk Assessment}
\nomenclature{TTA}{Time To Alert}
\nomenclature{TLOF}{Touch-down and Lift-off Area}
\nomenclature{UAM}{Urban Air Mobility}
\nomenclature{UAS}{Unmanned Aerial System}
\nomenclature{UAV}{Unmanned Aerial Vehicle}
\nomenclature{U-GBAS}{Urban GBAS}
\nomenclature{VAL}{Vertical Alert Limit}
\nomenclature{VPE}{Vertical Position Error}
\nomenclature{VTOL}{Vertical Take-Off and Landing}
\nomenclature{WTSA}{Wingtip-To-Safety Area}
\printnomenclature
\section{Introduction}\label{sec:introduction}
In the last years, the use of Unmanned Aerial Systems (UAS) has been growing at a very fast rate thanks to the development of lower cost technologies and the vast number of possible new applications that have appeared in the market. Numerous global companies are investing in the possibility of using small and medium sized Unmanned Aerial Vehicles (UAVs) for different services including cargo, surveillance, civil protection or delivery tasks, but also even more demanding use cases like personal mobility services (e.g. air taxis). It is therefore likely that in a near future a considerable amount of unmanned (and/or autonomous) aerial vehicles will need to share the airspace, particularly at low altitudes. The integration of UAS into a common air space with manned aircraft presents many challenges from the Air Traffic Management (ATM) point of view due to the potential different operational constraints. 

Since 2013 the Joint Authorities for Rulemaking on Unmanned Systems (JARUS) is getting importance in the development of guidelines for the integration of UAS into the airspace. JARUS is a group of experts from the National Aviation Authorities (NAAs) and regional aviation safety organizations. Its purpose is to recommend a single set of technical, safety and operational requirements for the certification and safe integration of UAS. Within this scope, published guidance and regulations include the safety assessment of Remotely Piloted Aircraft Systems (RPAS)~\citep{AMC_RPAS.1309} and Specific Operations Risk Assessment (SORA)~\citep{JARUS_SORA}. 

When considering the use of smaller UAS at lower altitudes and nearby or in urban environments, the design of both the Unmanned Air Traffic Management and the different on-board subsystems presents many new challenges. Moreover, one important goal is the design of these systems to guarantee a safe operation in potentially crowded urban environments, where collisions with other UAS as well as obstacles or people can be considered a major risk~\cite{Torens2021_HUAM}. Common agreed challenges regarding the UAS operations include Detect and Avoid (DAA), Command and Control (C2) and surveillance, among others. The integration of UAS in non-segregated airspace will also require for the definition of minimum navigation, communication and surveillance performance standards, as specified by the current development of UAS Concept of operations in Europe~\citep{EASA_Conops-Drones}. Of particular importance is the limitation of current approaches to provide a thorough analysis of the accuracy, continuity, availability and integrity of the navigation performance of UAVs in cities as compared to the current avionics Minimum Operational Performance Standards, such as~\citep{RTCA_DO-229D}. The major challenges in terms of navigation appear around four main aspects: 
\begin{enumerate}
\item the operation within an unrestricted air and ground space, 
\item the operation in challenging environment for sensors (e.g., Global Navigation Satellite System (GNSS) signal blockage and multipath) and near other (aerial) vehicles or obstacles, 
\item the need to use cost-efficient, small and light-weight sensors and
\item the necessity to consider very heterogeneous onboard-unit designs and requirements depending on the area of operation.
\end{enumerate}
In~\citep{CORUS}, a first assessment related to the threats and events that would lead to a technical and mechanical failure is provided. In relation to the navigation payload, the identified events are sensor or camera failures with respect to computer vision or a directional loss that can be caused by a Global Positioning System (GPS) failure due to GPS perturbations, compass failure, Inertial Measurement Unit (IMU) failure or altitude sensor failure. Although this description covers in general the main failure events, it does not provide an in-depth analysis of the specific threats to the navigation sensors, their implication in the navigation solution and possible mitigation measures. The complete safety assessment of Communication, Navigation and Surveillance (CNS) systems in Urban Air Mobility remains therefore an intensive research area\cite{Torens2021_HUAM}.

In this work, we first review the current regulation context for urban air mobility and derive new navigation requirements for drones operation, with focus on vertiport precision approach.
We then propose a multisensor navigation architecture based on the strength of different sensor technologies and services. The subsystems in the proposed architecture are discussed with focus on faults identification and challenges. Finally, initial validation results based on flight trials are provided, followed by outlook and conclusions.
\section{Urban Air Mobility}
For the safe operation of drones in the European air space, the EU has defined three categories to classify drone operations with a risk-based approach~\cite{EASA_EAR2022}: the {\em open}, {\em specific} and {\em certified} categories.
\begin{itemize}
\item \textbf{Open Category}: The operational risk in this category is considered low, no operational authorization is required and the safety is ensured by the civil drone operator complying with the requirements of the intended operation.
\item \textbf{Specific Category}: This category covers riskier operations that require a specific operational authorization where the drone operator is responsible to carry out a risk assessment. For this category, JARUS has developed a SORA methodology to facilitate the authorization and risk-assessment process~\cite{JARUS_SORA}.
\item \textbf{Certified Category}: In this category, the risk is assumed to be considerably higher and the certification of drone operator and vehicle is required to ensure safety. 
\end{itemize}
For the specific category, the JARUS SORA defines different Safety Assurance and Integrity Levels (SAIL) based on the intrinsic ground and air related risk of the UAS operation. For the different SAIL levels, the operator is required to show compliance with specific Operational Safety Objectives (OSO) with a certain robustness level (safety integrity and assurance) depending on the assigned SAIL for the operation.
Please note that {\em SORA integrity} refers to the safety gain of a specific OSO, whereas {\em navigation integrity} is a measure of trust that can be placed in the correctness of the information provided by the navigation system.
\subsection{Current Standardization Actions}\label{subsec:standards}
Currently, no navigation related standard is suitable for the operation of drones in any of the categories needing authorization or certification. For higher risk operations, which might cover the SAIL V-VI levels and the certified category, RTCA SC-228 has elaborated a navigation gap analysis focused on fixed wing aircraft operating in and out of traditional airports~\cite{RTCA_DO-397}. One major aspect is the lack of specific navigation requirements to support UAS operations, which is the case for both the specific and certified categories. GNSS is considered to be a primary system for most UAS operations. There is however, even for GBAS (Ground-based Augmenttion System) Approach Service Type (GAST) D, which is designed for autoland operations in Category III airports, no available standard that covers a fully autonomous landing below 12 feet. The operation within cities (i.e., non-restricted air-space) will make the standards deviate even more from an intented operation like landing in a vertiport. RTCA SC-228 plans to analyse UAS taxing operations, which will in the future help to provide guidelines for more challenging environmental conditions.
For low-risk operations (i.e., SAIL I-II), the European Organization for Civil Aviation Equipment (EUROCAE) has elaborated guideline material for the use of GNSS to support safety for these specific categories~\cite{Eurocae_ED-301}. These recommendations focus on the analysis of the OSO \#13 related to external services and therefore they provide recommendations to operators only for GNSS as a service. Plans exist to extend these guidelines to medium-risk operations. The complete navigation subsystem uses additional onboard sensors and systems and therefore other operational safety objectives must be analysed beyond the external services. 
%
%
\subsection{Navigation Requirements}
\label{sec:nav_req}
%
Table~\ref{tab:requirements} provides an overview of possible requirements for different use cases and safe operations in the specific category. The approach and landing requirements are derived from the previous vertiport analysis. The {\em enroute} requirements are extracted from \cite{EUSPA2023_userReportAviation} and are used here as a reference. The values in Table~\ref{tab:requirements} are possible guidance requirements which are under discussion and can change in the future.
Although SORA provides only a qualitative risk assessment methodology, for the derivation of the integrity risk requirement, we have assumed the following equation: 
\begin{align} 
    \textit{Integrity Risk (IR)} < \frac{10^{-(SAIL + 1)}}{\textit{Operation}}, \label{eq:IR}
\end{align}
where $SAIL$ refers to the SAIL risk level. \eq{eq:IR} is an adaptation of the assumption in \cite{EUSPA2023_userReportAviation} where we consider an operation can be either a flight hour or an approach and landing. For the achievement of other SORA integrity and assurance compliance, specific levels could also be linked to other quantitative industry standards. For instance, a Design Assurance Level (DAL) C for SAIL IV.
\begin{table*}[ht]
    \centering
    \begin{tabular}{p{2cm}|p{2.4cm}|p{1.8cm}|p{2cm}|c|c|c}
    \toprule
        \textbf{Operation}& \textbf{Accuracy} & \centering\textbf{Integrity Risk} & \textbf{HAL, VAL} & \textbf{TTA} & \textbf{Continuity} & \textbf{Availability} \\
         & (95\%)[m] & & \centering [m] & [s]& & [\%] \\
        \hline
        \midrule
        \multirow{4}{2cm}{\centering Enroute$^*$ (SAIL III)} & \multirow{8}{2cm}{HPE: 3 - 8, VPE: 4 - 13} & \multirow{4}{*}{1 - 1E-4/h} & \multirow{8}{2.2cm}{HAL: 25-27 (fixed wing), 10-14 (rotary);
        
        \vspace{1ex} VAL: 12-22 (fixed wing), 7-23 (rotary)} & \multirow{8}{*}{1 - 3} & \multirow{8}{*}{1 - 1E-4/h} & \multirow{8}{*}{99.99} \\
         & & & & & &\\
         & & & & & &\\
         & & & & & &\\
         \hhline{|-|~|-}
        \multirow{4}{2cm}{\centering Enroute$^*$ (SAIL IV)} &  & \multirow{4}{*}{1 - 1E-5/h} & & &  & \\
         & & & & & &\\
         & & & & & &\\
         & & & & & &\\
        \hline
        \centering Precision Approach (SAIL V - Certified) & HPE = 1.4-3.08, VPE: 1.22-2.66  & 1 - 1E-6/op to 1 - 1E-7/op & HAL: 3.93-8.2; VAL: 2.98-7.1 & $<$ 3 & 1 - 1E-8/op & $>$ 99.99\\
        \hline
        \hline
    \end{tabular}
    \caption{Navigation requirements for UAM flight phases. $^*$ Extracted from \cite{EUSPA2023_userReportAviation}.}
    \label{tab:requirements}
\end{table*}

\subsubsection{Vertiport Use Case}
\begin{figure}[tb]
    \centering
    \includegraphics[width=\linewidth]{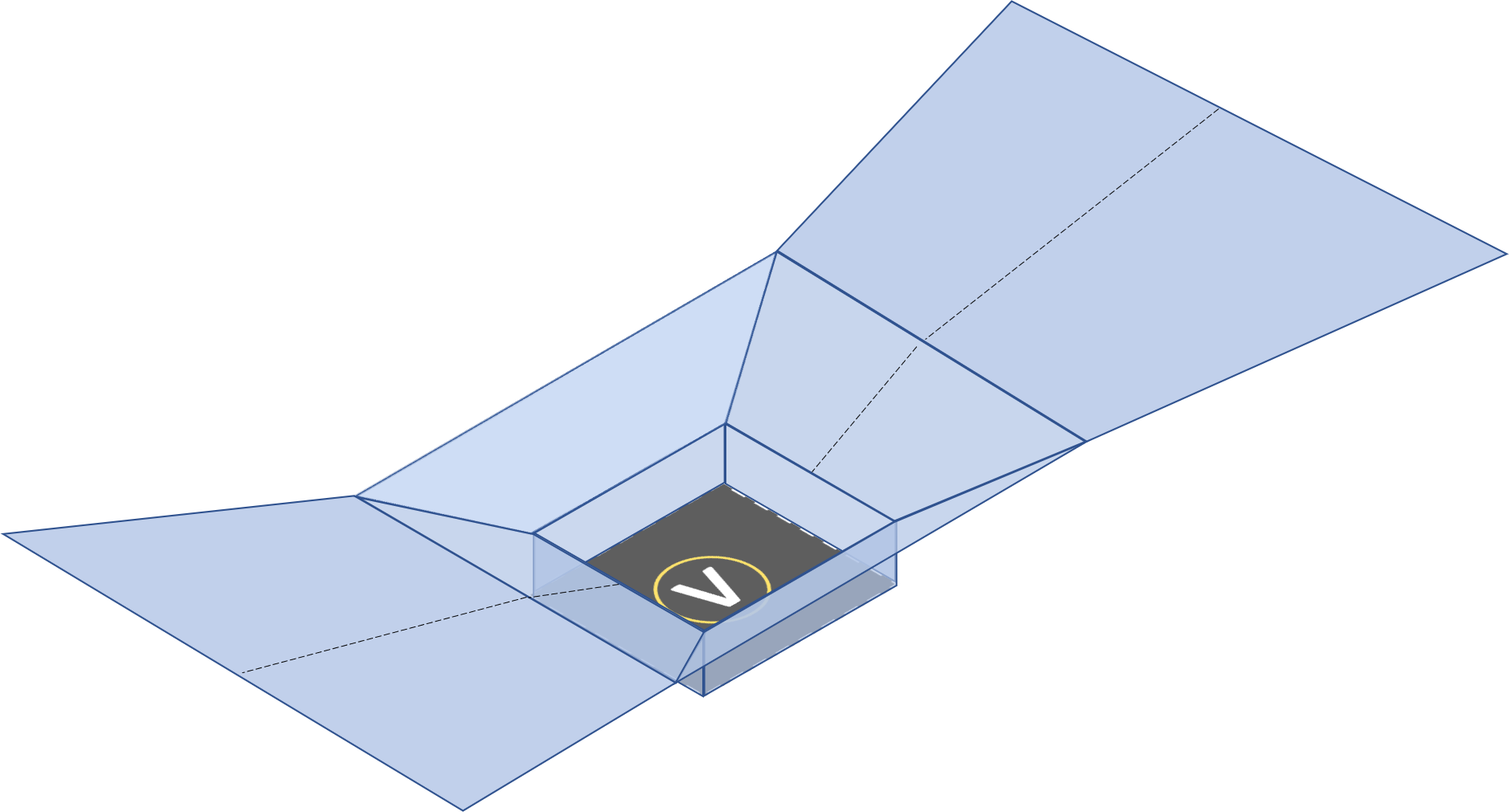}
    \caption{Example of a funnel-shaped vertiport design.}
    \label{fig:vertiport}
\end{figure}
\begin{figure}
    \centering
\includegraphics[width=0.8\linewidth]{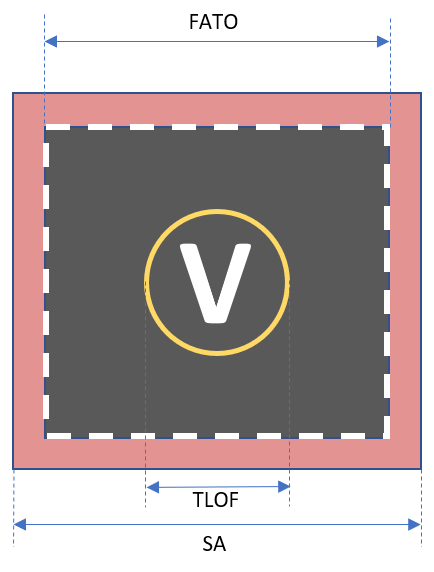}
    \caption{Vertiport areas definitions.}
    \label{fig:vertiport2D}
\end{figure}

For the final approach at the vertiport, we can first consider the potential associated horizontal requirements. The European Union Aviation Safety Agency (EASA) defines the {\em D-value} as the diameter of the smallest circle enclosing the Vertical Take-Off and Landing (VTOL) aircraft projection on a horizontal plane, while the aircraft is in the take-off or landing configuration ~\cite{EASA_PTS-VPT-DSN_2022}. We assume that it is expected that the vehicle dimension (D-value) must be inside the final approach and take-off area (FATO) with a high probability when reaching the hovering altitude. We will, therefore, use it as a criterion to derive navigation requirements, similar to an aircraft reaching the touchdown area during precision approaches to airport runways. The actual Touch-down and Lift-off (TLOF) area for vertiports is reached by the vehicle from the FATO hovering position, and it can be either situated inside the FATO footprint or could potentially be placed at the stand location and reached while hovering~\cite{AC_139.V-01}. This final touchdown manoeuvre is performed in a much controlled and low dynamics situation and can be supported also by visual references or additional proximity sensors and therefore we won't cover it in this paper. Figure~\ref{fig:vertiport2D} shows exemplary the different vertiport areas, including the safety area (SA).
The FATO area is proposed by EASA as twice the D-value as a reference, with a minimum value of the maximum between 1.5 times the D-value and the length of the rejected take-off distance (RTODV) for the required take-off procedure of the VTOL capable aircraft. The minimum FATO imposes tighter requirements for the navigation system and therefore we can consider:
\begin{equation}
    FATO = \max\{1.5D_\text{max}, RTODV_\text{max}\},
\end{equation}
where $D_\text{max}$ and $RTODV_\text{max}$ are the highest D-value and RTODV among the VTOLs that are expected to operate in that particular vertiport, since the vertiport would be designed with that criteria. Information about RTODV is not available at the moment, we will thefore focus on the design constraint based on the D-value.

For the horizontal nominal case, we can assume that the vehicle must be laterally inside the FATO area when it reaches the hovering altitude from the approach slope. The maximum allowed radial total system error (TSE) for the most constraining vehicle (that is, the largest one that drived the design of the vertiport) can be therefore expressed as a function of the margin between the wingtip and the end of the FATO area WTSA (Wingtip-To-Safety Area):
\begin{equation}
    WTSA = (FATO - D)/2.
\end{equation}
This margin represents the maximum allowed deviation from the center of the FATO or approach slope that could result in a collision with surrounding obstacles. Please note, that we assume conservatively here that obstacle collision can happen immediately beyond the FATO area, leaving the vertiport safety area (SA) as an additional buffer.
According to the vertiport design, the minimum WTSA would be:
\begin{equation}
    WTSA_\text{min} = (FATO - D_\text{max})/2 = D_\text{max}/4.
\end{equation}
Assuming a certain risk probability of finishing the approach outside the FATO area ($P_\text{out}$), we can express the equivalent TSE standard deviation as:
\begin{equation}
    \sigma_\text{TSE} = \frac{WTSA_\text{min}}{\sqrt{2}\cdot \text{erfc}^{-1}(P_\text{out})}, \label{eq:TSE}
\end{equation}
where $\text{erfc}$ is the complementary error function. In \eq{eq:TSE}, we have assumed the error to be Gaussian distributed or overbounded by a Gaussian distribution.
The TSE is the sum of the flight technical error (FTE), the navigation system error (NSE) and the path definition error (PDE) and can be related by their variances (again assuming Gaussian distribution or bounds) as:
 \begin{equation}
     \sigma^2_\text{TSE} = \sigma^2_\text{FTE} + \sigma^2_\text{NSE} + \sigma^2_\text{PDE}, 
 \end{equation}
where the latter is normally neglected. The NSE variance can be therefore written as:
\begin{equation}
    \sigma^2_\text{NSE} = \sqrt{\left(\frac{WTSA_\text{min}}{\sqrt{2}\cdot \text{erfc}^{-1}(P_\text{out})}\right)^2 - \sigma^2_\text{FTE}}.\label{eq:NSE}
\end{equation}
The FTE highly depends on the technical specifications of the vehicles, purpose, type and size. The analysis of the FTE is out of the scope of this work and we will assume for the moment a typical accuracy value of 0.5 m (i.e. at 95\%). We will also assume that the safety probability of violating the FATO area in the final approach ($P_\text{out}$) is not more stringent than the one of CAT-III landing, where a maximum allowed probability of 1E-6 is reported of exceeding the lateral limits of the landing box~\cite{FAA-AC120-28D}.
Finally, reported VTOLs prototypes of industry have a D-value ranging from 5.61 meters (eHang) up to 15.24 m (Uber)~\cite{Preis2022_vertiports}. It is unrealistic that vertiports will be designed only with the goal of supporting the smallest vehicles. If we assume a vertiport design based on the largest reported VTOL, the corresponding NSE standard deviation from \eq{eq:NSE} would be $\sigma_\text{NSE} \approx 0.74$ m.
An associated alert limit for a SAIL VI or certified operation with integrity risk 1E-7/h would be $HAL \approx 3.93$ m. If instead of the minimum value of the design of FATO we used the typical value of $2D$ mentioned in \cite{EASA_PTS-VPT-DSN_2022}. The horizontal accuracy (95\%) and alert limit would be 1.54 m and 8.2 m, respectively.

Related to vertical dimension requirements, there are two contributing aspects: the contribution of the vertical positioning in arriving short or far to the FATO in the longitudinal direction and the requirements related to the distance to the ground. For the longitudinal arrival to the FATO area at the end of the approach phase, the vertical error will have an important impact. Assuming the same probability of arriving short or late for a square FATO area, the vertical error can be simply obtained as a function of the approach/departure slope angle (also assuming a similar flight technical error for the vertical control):
\begin{equation}
    \sigma_\text{NSE,v} = \sigma_\text{NSE}\cdot \tan(\theta_\text{ADS}),
\end{equation}
where $\sigma_\text{NSE,v}$ is the standard deviation of the vertical navigation system error and $\theta_\text{ADS}$ is the Approach/Departure slope angle. In order to determine the slope angle, we use the reference volume type 1 in~\cite{EASA_PTS-VPT-DSN_2022}. For this volume, we can compute the slope based on the sizing of the high hover height take-off (TO) area located at 30.5 meters specified with a size of $4D$ and the FATO area at low hovering height of 3 m, which is in this case $2D$. The slope angle is therefore:
\begin{equation}
    \theta_\text{ADS} = \tan^{-1}\left(\frac{30.5-3}{2D-D}\right).
\end{equation}
For the maximum D-value of 15.24 m, the slope angle is therefore 61 deg. This leads to an standard deviation of 1.33 m and an associated vertical alert limit (at an integrity risk of 1E-7) of 7.1 m. 
The other possible vertical requirement for the reference volume 1 is related to the low hover height of 3m. If we assume that the vertical error cannot be larger than 3m for the same probability of 1E-6 assumed up to now, the associated vertical navigation system accuracy (95\%) would be 1.12 m, and an associated alert limit in the nominal case of 2.98 m.
%
%
\subsection{Navigation Technologies}\label{subsec:technologies}
GNSSs have been for decades the backbone of aerial vehicles navigation systems because of their wide availability. However, they are known to be unreliable in the urban environment and are relatively easy to be jammed or spoofed. Additionally, standalone GNSS vertical accuracy may not be sufficient for certain Urban Air Mobility (UAM) operations such as landing. 
The augmentation of GNSS with aircraft-based (ABAS), ground-based (GBAS) or satellite-based (SBAS) augmentation systems can increase accuracy and ensure safety. 
%
\begin{table*}
    \centering
    \def\arraystretch{1.3}
    \begin{tabular}{c|c|c|c|c|c|c}
    \toprule
        \textbf{\shortstack{Navigation \\ Technologies}} & \textbf{\shortstack{Enroute \\ Lateral}} & \textbf{\shortstack{Enroute \\ Vertical}} & \textbf{\shortstack{Approach \\ Lateral}}  & \textbf{\shortstack{Approach \\ Vertical}}  & \textbf{\shortstack{Vertiport \\ Lateral}}  & \textbf{\shortstack{Vertiport \\ Vertical}} \\
        \hline \midrule
        GNSS ARAIM & X & X & (X) & (X) & & \\ \hline
        SBAS & X & X & X & X & & \\ \hline
        GBAS & & & X & X & X & X\\ \hline
        GNSS/INS & X & (X) & X & (X) & X & (X) \\ \hline
        Vision & & & X & (X) & X & (X) \\ \hline
        \makecell{Barometer w/\\ Weather-correction} & & X & & (X) &  &\\\hline
        \makecell{Barometer w/\\ Ground-correction} & & & & (X) & & X\\ \hline
    \end{tabular}
    \caption{Navigation technologies relevance for UAM operations. The mark “(X)” means the technology could potentially add some benefits or would need to be adapted or improved for that use case.}
    \label{tab:technologies}
\end{table*}
Current ABAS developments focus on proving that Advanced Receiver Autonomous Integrity Monitoring (ARAIM) is able to provide a robust operation for horizontal services and vertical guidance via the use of Multi-frequency and Multi-constellation GNSS. Although ARAIM can attain high levels of integrity, the achievable protection levels are limited by the overbounding of error distributions to account for worst case expected performances. Moreover, due to the required maneuvers in terminal vicinity and precision procedures, the availability and continuity of the system may be compromised by the loss of satellites or high presence of cycle slips~\citep{crespillo2022}. In this sense, the support of several various augmentation systems is essential to guarantee all the navigation requirements and extend the achievable accuracy and integrity for stringent operations, such as UAM ones (especially in vertiport nearness).
GBAS provides real-time differential corrections for GNSS to allow aircraft landings with high accuracy and integrity, but as GNSS-based system is susceptible to jamming and spoofing. 
A GNSS fusion with an Inertial Navigation System (INS) is effective for coasting during GNSS signals losses but does not enhance precision. This is especially true with respect to vertical positioning, because of the inherent instability of the INS vertical channel. INS can additionally provide attitude and thus heading information. 
In aviation, barometers are used for the computation of pressure altitude. This is a virtual altitude that, depending on weather conditions, can differ by up to hundred of meters from geodetic altitude, which is employed by GNSSs. The latter is also the altitude expected to be used for vertical navigation of future UAM applications~\citep{EASA_CARS, CORUS_Conops, ICARUS_D3_1}. The use of barometric altimeters for UAM navigation, potentially also in combination with GNSS, requires therefore an accurate correction of pressure altitude and a rigorous conversion to geodetic altitude.
Radar altimeters may be a valid alternative to barometric ones. They present several limitations though. Interpolation of digital terrain elevation data (DTED) based on horizontal location is needed to produce geodetic altitude measurements. Accurate horizontal positioning is therefore necessary, even though sharp terrain elevation variations may not be observable with available DTED. Furthermore, dynamic variations of the orientation of radar sensors with respect to the ground may lead to potentially large errors. Finally, measurement errors increase with increasing altitude above ground. Nevertheless, radar altimeters may be useful for vertiport operations of VTOL vehicles because of: approximately flat vertiport surface and knowledge of its elevation, negligible attitude dynamics, proximity to the ground, and absence of obstacles~\citep{Radalt2005, Radalt2019}. The aforementioned limitations may however not allow to effectively use radar altimeters for other flight phases.
Lidar (light detection and ranging) sensors represent a further option for vertical navigation. Their working principle is similar to radar altimeters, but it is based on the transmission, reflection and reception of laser pulses, rather than of radio waves. Lidar sensors can provide more accurate and higher resolution data than radars. However, the performance of lidar is more susceptible to weather conditions, in particular to fog and rain~\citep{Lidar2023}.
\begin{figure*}[t]
\includegraphics[width=\linewidth]{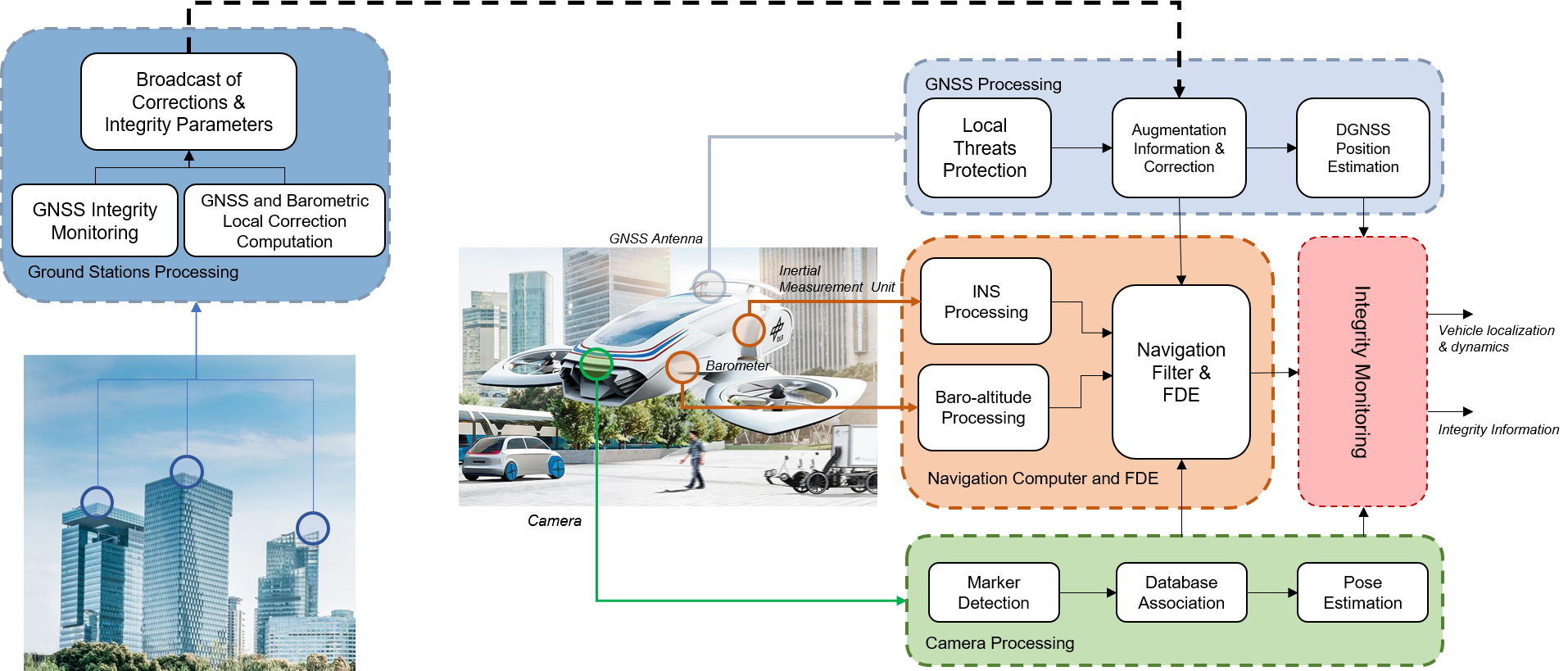}
\caption{Navigation architecture for vertiport operations.}
\label{fig:architecture}
\end{figure*}
Visual navigation has great potential for UAM applications. 
Digital images are captured using airborne cameras, and the images are processed onboard in real-time to search key features. For enroute flight, vision can be used for lateral positioning based on map-matching \cite{Stouffer2020_CNS_UAM}. If a known geo-referenced pattern in the database, such as a fiducial marker at the vertiport, is detectable in the camera view, the 6 degrees of freedom relative poses (positions and attitude) between the camera and the pattern can be estimated using the pixel points detected in the images. This enables the possibility to apply cameras as a complementary navigation sensor during the take-off and landing phases at vertiports. For low-light conditions (e.g., night operations), infrared cameras and lighting patterns at vertiports can be used as proposed in \cite{Veneruso2022}. 
The authors in \cite{RTCA_DO-397, Stouffer2020_CNS_UAM} have identified that performance standards and mature certifiable solutions of vision-based navigation for UAM/UAS application are still missing. In particular, it is challenging to quantify the integrity of visual navigation in a reliable way, as well as certify the solution. 
%
%
In our proposed architecture, we have considered INS to support the attitude determination. This solution may be complemented by magnetometers, in particular with respect to the heading. These sensors are however susceptible to electromagnetic interference, which could pose a threat to the attitude determination and may be difficult to consider within an integrity monitoring algorithm.
\section{A First Navigation Architecture Design for Vertiports Approach}\label{sec:architecture}
This section focuses on a potential architecture and safety design analysis for aerial vehicles approach and landing operations. The navigation system architecture considered in this paper makes use of the following sensors and technologies: Local Differential GNSS Augmentation (Urban GBAS), Inertial Measurement Unit (IMU), Barometers and Cameras. Each of these technologies can provide important relative or absolute information and would have a major role depending on the UAS operation. As an example, Table~\ref{tab:technologies} provides an overview of in which operation each technology may have a larger importance or contribution to the navigation solution.

A general system overview and design is presented in \figref{fig:architecture}.
\begin{figure*}
    \centering
    \begin{scaletikzpicturetowidth}{1.95*\columnwidth}
\begin{tikzpicture}
 [scale=\tikzscale,
    and/.style={and gate US,thick,draw,fill=red!60,rotate=90,
        anchor=east,xshift=-1mm},
    or/.style={or gate US,thick,draw,fill=gray!60,rotate=90,
        anchor=east,xshift=-1mm},
    be/.style={circle,thick,draw,fill=green!60,anchor=north,
        minimum width=0.7cm},
    label distance=3mm,  every label/.style={blue},
    event_up/.style={rectangle,thick,draw,fill=gray!10,text width=4cm, text centered,font=\sffamily,anchor=north},
    event/.style={rectangle,thick,draw,fill=gray!10,text width=2cm, text centered,font=\sffamily,anchor=north},
    edge from parent/.style={very thick,draw=black!70},
    edge from parent path={(\tikzparentnode.south) -- ++(0,-1.6cm)-| (\tikzchildnode.north)},
    level 1/.style={sibling distance=5cm,level distance=3cm, growth parent anchor=south,nodes=event},
    level 2/.style={sibling distance=3cm, level distance=2cm},
    level 3/.style={sibling distance=3cm},
    level 4/.style={sibling distance=3cm}
    ]
 \node (g1) [event_up]{Positioning Failure \\ ($P_\text{HMI}$)}
            child{node (e1) {Nominal Sensor Fusion failure \\ ($P_\text{H0}$)} 
            }
        child{node (e2) {GNSS residual fault \\ ($P_\text{GNSS}$)} 
        }
        child {node (e3) {Barometer residual fault \\ ($P_\text{baro}$)}
        }
        child {node (e4) {INS fault \\ ($P_\text{INS}$)}
        }
        child {node (e5) {Vision fault \\  ($P_\text{vis}$) } 
        };
   \node [or]   at (g1.south)   []  {};
\end{tikzpicture}
\end{scaletikzpicturetowidth}
    \caption{Navigation integrity tree.}
    \label{fig:integrityTree}
\end{figure*}
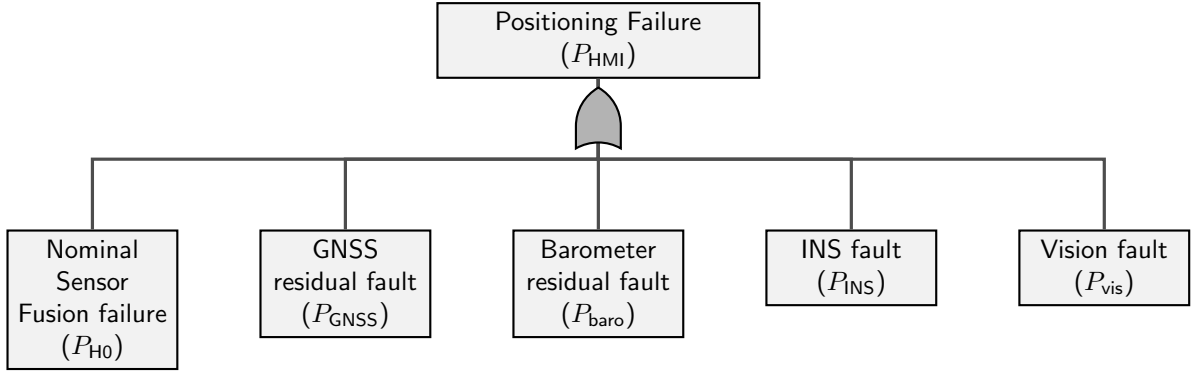
The system architecture consists of a ground reference infrastructure part and a multisensor onboard navigation unit.

The {\em ground reference infrastructure} consists of spatially distributed ground stations equipped with GNSS antennas and receivers and weather stations. The ground infrastructure perform the following tasks:
\begin{itemize}
\item Computes GNSS local corrections
\item Computes reference pressure levels
\item Performs GNSS integrity monitoring for the signal in space
\item Broadcast the corrections and the integrity information
\end{itemize}

The {\em multisensor onboard unit} consists of several sensors installed on the vehicle and different processing blocks:
\begin{itemize}
\item GNSS Processing performing dedicated local GNSS monitoring, reception of GNSS augmentation information and local corrections, and computation of a reference Differential GNSS position.
\item Camera Processing detecting landing pad markers to estimate a camera-based vehicle pose based on monocular camera.
\item The Multisensor Navigation Computer processes Inertial Measurement Unit (IMU) data into inertial navigation system quantities, converts onboard pressure to  barometric geodetic altitude and combines corrected GNSS, INS, Baro-altitude and Camera data to compute a position, velocity, attitude and timing solution of the system. This block also performs different fault monitoring levering the redundancy between the different sensor information.
\item The integrity monitoring processing evaluates the performance of the different estimations and fault monitors in the system to derive protection levels and associated integrity information on the final position, velocity and attitude solutions.
\end{itemize}
For the design of fault detection mechanisms and perform an assessment of the integrity of the system, it is essential to derive a threat model of the system. Figure~\ref{fig:integrityTree} provides a possible general overview of the different components that would contribute to a failure of the positioning system. They include the possible failure of the nominal sensor fusion algorithm (typically based on Kalman filtering), the residual GNSS and barometric altitude faults, the failure of the inertial system and the fault of the vision-based positioning determination. In the next, the different subsystems in \figref{fig:architecture} are further detailed.
\section{Subsystems  Considerations}\label{sec:subsystems}
\subsection{Global Navigation Satellite System (GNSS)}
In this section, we provide with further information about the subsystems of the multisensor navigation architecture, also commenting on the advantages and challenges of using these technologies.

\subsubsection{Airborne GNSS}
GNSS positioning in unrestricted and close-to-ground airspace can be significantly affected by various local threats that cannot be corrected or compensated for through augmentation information. Therefore, GNSS positioning algorithms must be safeguarded against the impact of these local effects, which can distort GNSS code and carrier measurements, rendering their errors inconsistent with the nominal error model assumed by the estimation algorithm. A list of pertinent local threats and possible detection techniques is provided in \cite{ITST2018} for land-based applications, which are also applicable to close-to-ground flying operations. Several protection measures can be considered, including:
\begin{itemize}
\item Radio-Frequency Interference (RFI) Detection Based on Power Spectral Density (PSD) Analysis: Commercial-off-the-shelf (COTS) GNSS receivers can provide information about the frequency spectrum or I/Q samples for computation. By comparing the nominal PSD with the currently received one, various types of narrow and wideband interference can be detected \citep{vennarini2020detection}.
\item Excessive Multipath Detection: For urban scenarios, a multipath detector can be developed based on the Code-minus-Carrier (CMC) observable, using the expected rate change of multipath as a criterion~\cite{Caamano2020}. If strong multipath is detected on a specific channel, measurements from that satellite can be discarded.
\item Data Editing: The practice of discarding measurements through reasonable checks and decisions has proven effective in reducing the presence of large, unbounded measurement errors \citep{GarciaCrespillo2022_architecture}. This process, known as data editing, may involve applying a CN0 mask, discarding measurements based on the loss of lock indicator (LLI), or removing measurements when both L1 and L2 measurements are unavailable, among other criteria.
\end{itemize}
Excessive protection against local threats can impact measurement availability. Therefore, it is essential to strike a balance between the nominal error model and protection against threats to meet availability requirements. However, it is expected that continuity and availability of the final solution will be ensured through the presence of additional sensors, such as inertial measurement units and their filter-based integration~\citep{GarciaCrespillo2022_architecture}. Consequently, it is crucial to guarantee measurement integrity even if this reduces the number of available (i.e., reliable) code and carrier phase measurements.
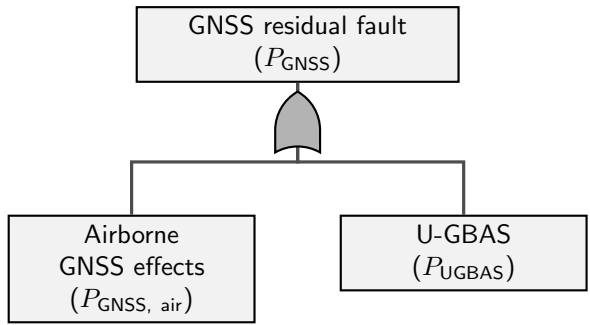
\begin{figure}[bt]
    \centering
    \begin{scaletikzpicturetowidth}{0.95\columnwidth}
\begin{tikzpicture}
 [scale=\tikzscale,
    and/.style={and gate US,thick,draw,fill=red!60,rotate=90,
        anchor=east,xshift=-1mm},
    or/.style={or gate US,thick,draw,fill=gray!60,rotate=90,
        anchor=east,xshift=-1mm},
    be/.style={circle,thick,draw,fill=green!60,anchor=north,
        minimum width=0.7cm},
    label distance=3mm,  every label/.style={blue},
    event_up/.style={rectangle,thick,draw,fill=gray!10,text width=4cm, text centered,font=\sffamily,anchor=north},
    event/.style={rectangle,thick,draw,fill=gray!10,text width=3cm, text centered,font=\sffamily,anchor=north},
    edge from parent/.style={very thick,draw=black!70},
    edge from parent path={(\tikzparentnode.south) -- ++(0,-1.2cm)-| (\tikzchildnode.north)},
    level 1/.style={sibling distance=5cm,level distance=2cm, growth parent anchor=south,nodes=event},
    level 2/.style={sibling distance=3cm, level distance=2cm},
    level 3/.style={sibling distance=3cm},
    level 4/.style={sibling distance=3cm}
    ]
 \node (g1) [event_up]{GNSS residual fault \\ ($P_\text{GNSS}$)}
            child{node (e1) {Airborne GNSS effects\\ ($P_\text{GNSS, air}$)}
            }
        child{node (e2) {U-GBAS \\ ($P_\text{UGBAS}$)}
        };
 
   \node [or]   at (g1.south)   []  {};
\end{tikzpicture}
\end{scaletikzpicturetowidth}
    \caption{Fault tree of GNSS measurements (separated between airborne and differential).}
    \label{fig:gnss_fault_tree}
\end{figure}
\subsubsection{Differential GNSS (U-GBAS)}
Classical GBAS was developed with the requirement to replace established landing systems, in particular ILS (Instrument Landing System). As a result, many design decisions were made with established operational aspects and the need to ensure compatibility with existing systems in mind. The existing standardised systems are currently limited to the use of GPS and single frequency measurements. Although there are ongoing efforts to take advantage of developments in satellite navigation by integrating additional constellations and a second frequency, backward compatibility still limits the freedom to change design philosophies.
In urban air mobility, such an ecosystem of constraints does not yet exist, providing more room for innovative ideas and concepts. New data links, the use of additional GNSS constellations as well as new signals can compensate for some of the shortcomings and problems encountered in an urban environment.
Our proposed concept \cite{gerb2023} of Urban GBAS (U-GBAS) involves the deployment of a network of local reference receivers mainly at vertiports within a given service area. These reference receivers form a network covering an urban or metropolitan area, providing accurate and reliable local GNSS corrections for UAM operations, particularly during take-off and landing. Additionally, satellite signals are monitored for faults and larger distortions, be it from various possible hardware failures aboard the satellites or from ionospheric effects along the signal path \cite{lee2023}. This can reduce user (airborne) monitoring requirements and therefore improve system availability especially for users with lower-grade GNSS hardware on board. Furthermore, especially at the critical take-off and landing sites, installed ground receivers provide a possibility to monitor for local interference, another potential threat to GNSS with its weak signals.
The core service volume of the U-GBAS system includes the polygon spanned by the reference sites. Unlike the classical GBAS system, which relies on expensive antennas installed in open sky free of obstacles, the proposed U-GBAS system envisions the use of smaller and cheaper hardware, including antennas, while still covering all relevant civil frequency bands and multiple constellations. This approach is intended to make it easier and less costly to integrate U-GBAS into a future urban airspace.
\subsection{GNSS/INS}
The Inertial Navigation System (INS) is typically in multisensor solutions the central sensor that leads the high frequency positioning solution. The rest of the sensors provide information that is used to correct the drift of the INS. The inertial measurement unit is not affected by the environment or scenario (e.g., loss of satellite signals due to urban canyon or bad image quality). 
The IMU measurements (i.e., specific forces and angular rates) are typically processed by a strapdown computer in order to obtain the INS attitude, position and velocity over time. The solution of the INS is normally the main positioning solution provided by the integrated system. In order to prevent the estimated pose from drifting over time due to the IMU error processes, an error state Extended Kalman filter is implemented to calibrate the INS system over time thanks to GNSS or other information. An error state version of the EKF is normally chosen so that effects due to linearization of the INS differential equations does not have a significant impact on the representations of the error estimation provided by the filter.
\subsection{Barometer-based Vertical Navigation Subsystem}\label{subsec:baro_theory}
Barometers are traditionally employed in aviation for the computation of altitude information, known as pressure altitude, based on the International Standard Atmosphere (ISA) model~\citep{ICAO7488}. This model is used to obtain the altitude, above a certain isobar reference, that corresponds to a measured air pressure. Above so-called transition altitudes, the constant temperature and pressure of the ISA mean sea level (MSL) standard isobar are used to obtain the so\textendash called QNE pressure altitude, which is used for relative vertical separation of aircraft. Depending on weather conditions, QNE pressure altitude can differ by up to hundreds of meters from geodetic altitude, which is employed by GNSS~\cite{Baro2021}. The latter is also the altitude expected to be used for vertical navigation of future UAM applications~\citep{EASA_CARS, CORUS_Conops, ICARUS_D3_1}. 
%
\begin{figure}[tb]
    \centering
    \begin{scaletikzpicturetowidth}{0.95\columnwidth}
\begin{tikzpicture}
 [scale=\tikzscale,
    and/.style={and gate US,thick,draw,fill=red!60,rotate=90,
        anchor=east,xshift=-1mm},
    or/.style={or gate US,thick,draw,fill=gray!60,rotate=90,
        anchor=east,xshift=-1mm},
    be/.style={circle,thick,draw,fill=green!60,anchor=north,
        minimum width=0.7cm},
    label distance=3mm,  every label/.style={blue},
    event_up/.style={rectangle,thick,draw,fill=gray!10,text width=4cm, text centered,font=\sffamily,anchor=north},
    event/.style={rectangle,thick,draw,fill=gray!10,text width=3cm, text centered,font=\sffamily,anchor=north},
    edge from parent/.style={very thick,draw=black!70},
    edge from parent path={(\tikzparentnode.south) -- ++(0,-1.2cm)-| (\tikzchildnode.north)},
    level 1/.style={sibling distance=5cm,level distance=2cm, growth parent anchor=south,nodes=event},
    level 2/.style={sibling distance=3cm, level distance=2cm},
    level 3/.style={sibling distance=3cm},
    level 4/.style={sibling distance=3cm}
    ]
 \node (g1) [event_up]{Barometric geodetic altitude failure \\ ($P_\text{baro} = 1.56\textrm{E-4}/\textrm{h}$)}
            child{node (e1) {Ground weather information failure \\ ($P_\text{baro, gnd} = 6.7\textrm{E-5}/\textrm{h}$)}
            }
        child{node (e2) {Airborne static pressure sensing failure \\ ($P_\text{baro, air} =6.9\textrm{E-5}/\textrm{h}$)  }
        };
 
   \node [or]   at (g1.south)   []  {};
\end{tikzpicture}
\end{scaletikzpicturetowidth}
    \caption{Fault tree of the ground\textendash corrected barometric geodetic altitude.}
    \label{fig:baro_fault_tree}
\end{figure}
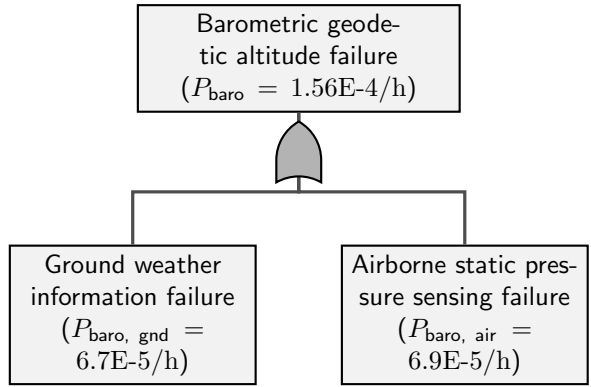
For UAM vertical navigation during approach and operations in vertiport vicinity, we envision a correction of pressure altitude based on pressure and temperature measurements performed at the vertiport. This corresponds to using as a reference an estimate of the isobar surface that is currently aligned with the vertiport surface. We refer to this approach as the \textit{ground\textendash correction} approach. This is then followed by a conversion procedure to geodetic altitude that takes into account the differences in terms of altitude scale and reference that exist between (corrected) pressure altitude and geodetic altitude~\citep{Baro2021}.

The ground-correction approach is similar to but different from other approaches that in aviation are used for correcting pressure altitude with the aim of ensuring separation from the ground in airport vicinity operations. These corrections consist in setting as reference the isobar at the airport’s location (QFE approach), or an estimate of the actual MSL isobar (QNH approach) at the same horizontal location of the airport~\citep{RTCA_DO-384}. For these corrections, only pressure settings are considered. Indeed, the reference temperature employed in QNH, QFE or QFE pressure altitude computation is the constant ISA MSL temperature. Pressure altitude above a certain reference overestimates the true geopotential altitude above that reference when the actual atmospheric temperature is lower than predicted by the ISA, which is of course safety\textendash critical. Airplanes receive QFE or QNH pressure settings from airports. Depending on the vertical and horizontal distance from the transmitting airport, the QNH\textendash { or} QFE\textendash corrected pressure altitudes can more or less deviate from the actual geopotential altitude above MSL or above the airport, respectively. Besides, the QNH\textendash or QFE\textendash pressure settings are rounded down to the nearest lower whole hPa~\citep{ICAOMeteo}, with therefore a maximum rounding error of of 99 Pa. This rounding error corresponds to a pressure altitude offset of more than 8 m, at ISA MSL conditions.
Finally, it can be assumed that QFE or QNH corrections degrade over time, as both are contained in half\textendash hourly or hourly weather reports transmitted by airports~\citep{ICAOMeteo}.

Unlike for the QFE or QNH and as mentioned above, in our ground\textendash correction approach we propose to directly employ ground temperature measurements too in order to mitigate the temperature effects. Additionally, for the ground\textendash approach we propose a pseudo\textendash continuous transmission of the ground weather measurements (i.e. whenever new measurements are available), whereby the pressure readings are rounded not more than to the closest Pa.

Ground\textendash correction may also be based on weather measurements performed on top of buildings for the transition between flight phases at higher altitudes and approach/landing. 

The usage of weather measurements from a vertiport for pressure altitude computation may be less suitable during UAM flight phases happening at relatively large distances from that vertiport. This may be the case during enroute operations. In those scenarios, the correction of pressure altitude may be performed based on estimates of local weather parameters obtained through the interpolation of weather data made available by an external provider~\citep{ERA5} on numerous vertical levels. We refer to this approach as the \textit{weather\textendash corrected pressure altitude}~\citep{Baro2021}. 


We call the geodetic altitude obtained from any corrected pressure altitude the \textit{barometric geodetic altitude}. A fault tree can be generated for the safety assessment of barometric geodetic altitude computation in UAM. The fault tree can be structured with two main top\textendash level branches, as shown in \figref{fig:baro_fault_tree}. One branch\textemdash here the right one\textemdash is related to airborne pressure measuring. The left branch is related to the data used for the pressure altitude correction. 
In the ground\textendash correction approach the left branch shall consider the components of the system providing ground weather measurements. Here, we have directly included as fault rate the inverse of a typical mean time between failure (MTBF), i.e. 15000 hours, of suitable pressure and temperature transmitters~\citep{Vaisala}. In case the weather\textendash correction approach was considered, then the left branch would need to be related to the used weather dataset instead~\citep{Baro2022}.
The fault rate in the right branch of \figref{fig:baro_fault_tree} is derived from the analyses of~\cite{Lerro2021}, which focus on the safety assessment of an air data system suitable for UAM applications.

\begin{figure}[t]
    \centering
    \includegraphics[width=\linewidth]{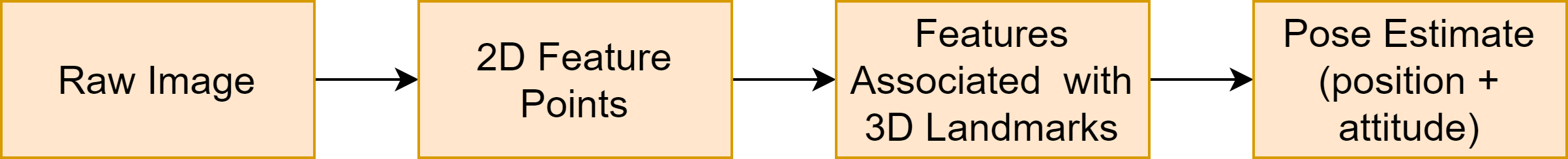}
    \caption{Visual positioning processing chain.}
    \label{fig:VisionCoreProcedure}
\end{figure}
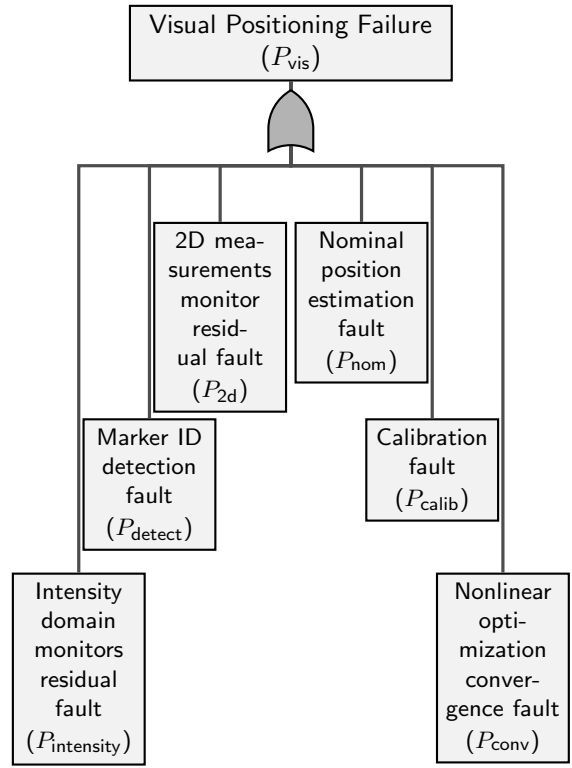
\begin{figure}[t]
    \centering
    \begin{scaletikzpicturetowidth}{0.95\columnwidth}
\begin{tikzpicture}
 [scale=\tikzscale,
    and/.style={and gate US,thick,draw,fill=red!60,rotate=90,
        anchor=east,xshift=-1mm},
    or/.style={or gate US,thick,draw,fill=gray!60,rotate=90,
        anchor=east,xshift=-1mm},
    be/.style={circle,thick,draw,fill=green!60,anchor=north,
        minimum width=0.7cm},
    label distance=3mm,  every label/.style={blue},
    event_up/.style={rectangle,thick,draw,fill=gray!10,text width=4cm, text centered,font=\sffamily,anchor=north},
    event/.style={rectangle,thick,draw,fill=gray!10,text width=1.5cm, text centered,font=\sffamily,anchor=north},
    edge from parent/.style={very thick,draw=black!70},
    edge from parent path={(\tikzparentnode.south) -- ++(0,-1.2cm)-| (\tikzchildnode.north)},
    level 1/.style={sibling distance=3cm,level distance=2cm, growth parent anchor=south,nodes=event},
    level 2/.style={sibling distance=3cm, level distance=2cm},
    level 3/.style={sibling distance=3cm},
    level 4/.style={sibling distance=3cm}
    ]
 \node (g1) [event_up]{Visual Positioning Failure \\ ($P_\text{vis}$)}
            child {node (e1) at (4.5,-5) {\small Intensity domain monitors residual fault \\ ($P_\text{intensity}$) }
            }
            child{node (e2) at (2.5,-2.8) {\small Marker ID detection fault \\ ($P_\text{detect}$) }
            }
            child {node (e3) at (0.5,0) {\small 2D measurements monitor residual fault \\ ($P_\text{2d}$) }
            }
            child{node (e4) at (-0.5,0) {\small Nominal position estimation fault \\ ($P_\text{nom}$)} 
            }
            child {node (e5) at (-2.5,-2.8) {\small Calibration fault \\ ($P_\text{calib}$) }
            }
            child {node (e6) at (-4.5,-5) {\small Nonlinear optimization convergence fault \\ ($P_\text{conv}$) }
        };
   \node [or]   at (g1.south)   []  {};
\end{tikzpicture}
\end{scaletikzpicturetowidth}
    \caption{Vision subsystem integrity tree.}
    \label{fig:VisionIntegrityTree}
\end{figure}
\subsection{Camera}
\begin{figure*}[bt]
    \centering
    \includegraphics[trim={1cm 1.5cm 0.5cm 1cm},clip,width=\linewidth]{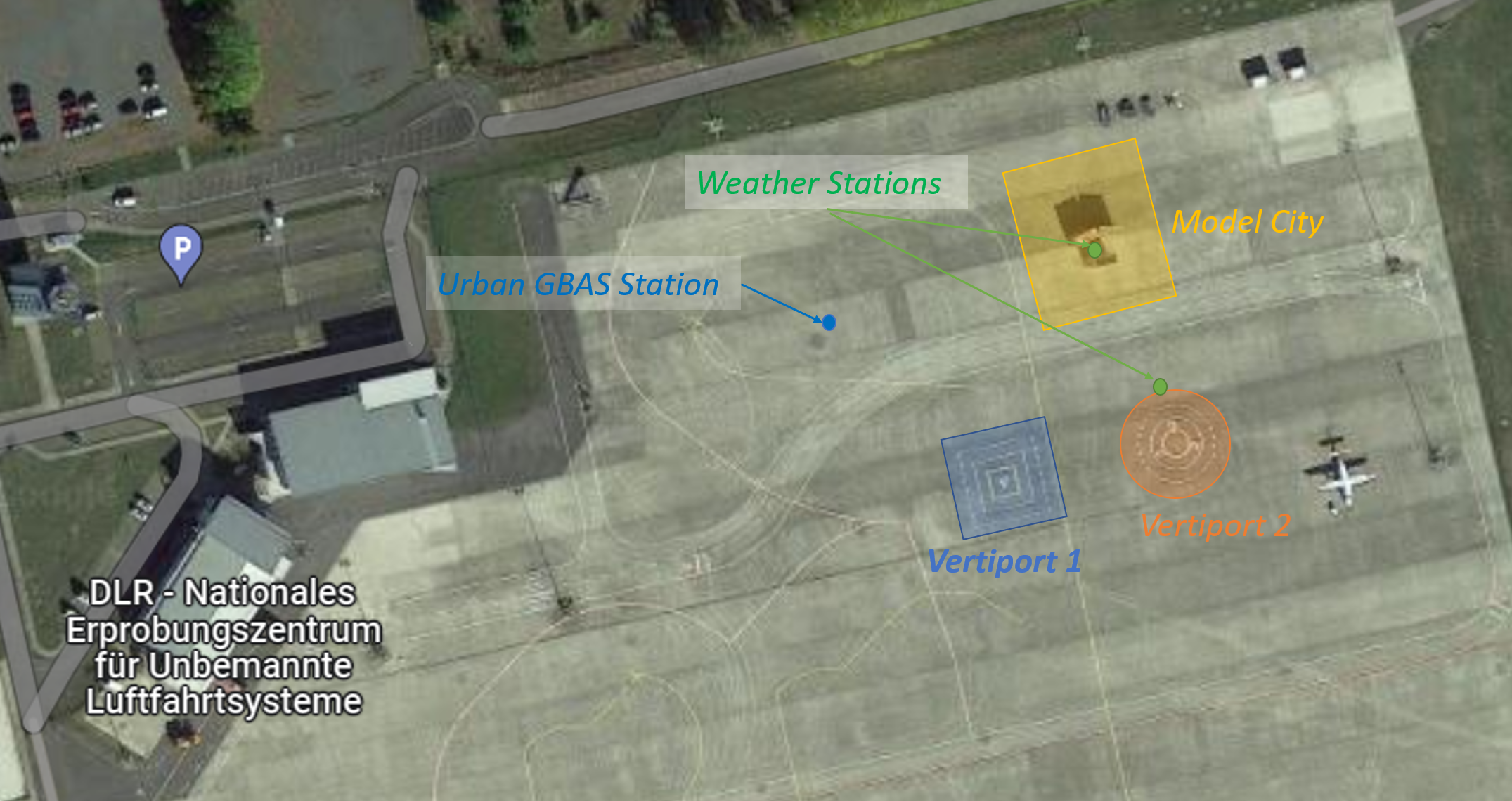}
    \caption{Test set-up for navigation systems during HorizonUAM project in Cochstedt airport, Germany.}
    \label{fig:cochstedt}
\end{figure*}
For UAM vertiport operations, machine vision can play an important role as a complement to radio-based navigation systems due to its high precision and availability in environments with radio interference. Nevertheless, for safety-critical UAM applications, the standardization and certification gaps for computer-vision-based navigation methods have limited the application of this technology. To the best of our knowledge, there is no standard yet to certify visual navigation methods for UAM applications. In guidance material published by RTCA to identify navigation gaps for unmanned aircraft systems \cite{RTCA_DO-397}, it was noted that certification of the vision systems will be difficult and there is a gap in standardized methods to assess the integrity performance. Identifying the critical fault modes to monitor and quantifying the algorithm integrity are the main challenges of certifying visual navigation systems. For visual navigation, the fault occurrence frequency and the consequent error magnitude are highly dependent on the scenarios, the onboard sensor setup and the specific processing algorithm applied. These aspects are taken into account in our research and design to take a step towards future certifiable systems.


Depending on the application scenario, different categories of visual navigation methods (from visual odometry for relative motion estimation to deep-learning-based map matching, see \cite{Zhu2022} for a more detailed review) can be used to fulfill the design need. For safety-critical UAM landing, high integrity is required when the vision subsystem is used for navigation purposes. As a result, the design of the visual navigation method should satisfy the following criteria to achieve high integrity:
\begin{itemize}
    \item The processing procedure must be interpretable so that the residual error can be properly quantified and monitored.
    \item The algorithm should be robust to lighting condition changes.
    \item There must be measurement redundancy for integrity checking.
\end{itemize}
According to the above criteria, we choose to apply a visual positioning algorithm using the feature points of the georeferenced markers. The chosen algorithm also has the advantage that the position of the onboard camera can be estimated independently at each time snapshot, so it is not affected by the error correlation over time, which is challenging to model conservatively. From the implementation and operation points of view, the cost and effort of painting and georeferencing markers on vertiports is usually quite affordable. 

The core procedure of positioning the camera includes processing in different data spaces, as shown in \figref{fig:VisionCoreProcedure}.

Faults that can result in positioning integrity risk or continuity risk may occur at each individual phase and propagate in the processing. 
In order to quantify the visual positioning integrity, all the errors in different domains in the processing must be properly monitored. Therefore, we design the visual positioning algorithm for UAM landing by allocating the integrity budget to different processing blocks, following the multi-domain integrity monitoring framework for visual navigation proposed in \cite{Zhu2022}. 
%
%

As a subsystem providing additional availability and accuracy to the multisensor navigation solution during approach and landing phases, the vision system is used for navigation only if the integrity risk of the subsystem itself fulfills the integrity requirements. For our design for these specific application scenarios, the vision subsystem faults can be broken down as in \figref{fig:VisionIntegrityTree}. It should be mentioned that the decomposition is a preliminary proposal according to the subsystem requirements analysis, given that there is no standard yet to quantify the exact values of the integrity requirements on visual navigation systems. Such a decomposition of the integrity tree provides a first basis for the design and implementation of the integrity monitors for visual positioning in our research. Further developments and validations are necessary to obtain more precise requirements and practically achievable performance of the visual positioning system. 

%
\section{Initial Subsystems Proof of Concept}\label{sec:demo}
%
%
\begin{figure}[t]
    \centering    \includegraphics[width=\linewidth]{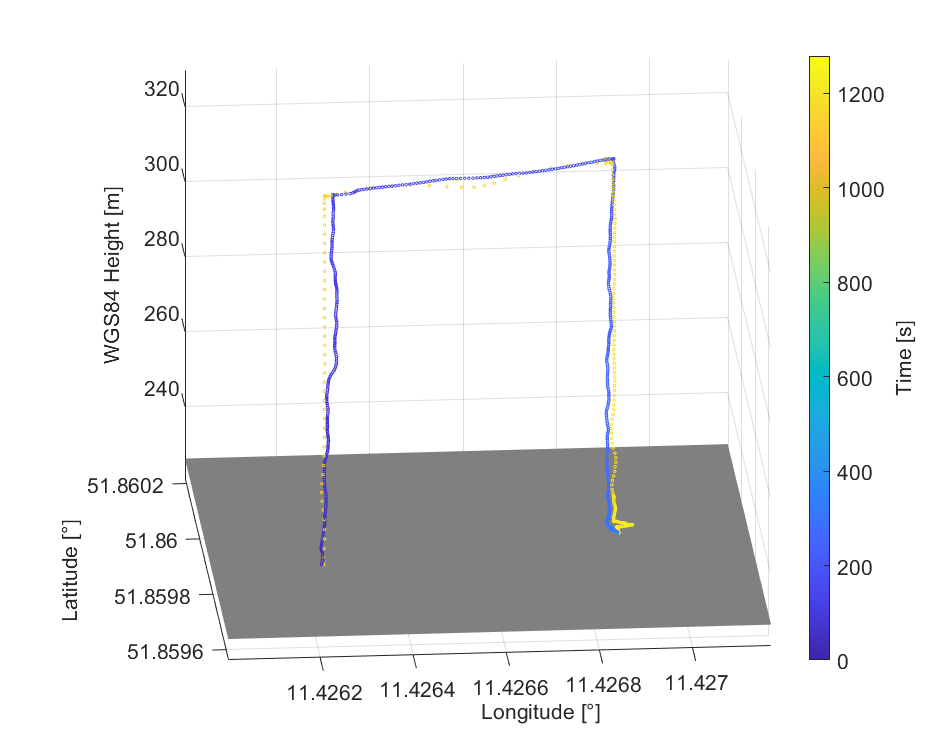}
    \caption{Reference trajectory of short flights between Vertiport 1 and 2 in \figref{fig:cochstedt}.}
    \label{fig:UGBAS_ref_trajectory}
\end{figure}
\begin{figure}[t]
    \centering
     \centering
     \setlength\figwidth{0.87\linewidth}
     \setlength\figheight{1.2\linewidth}
    \input{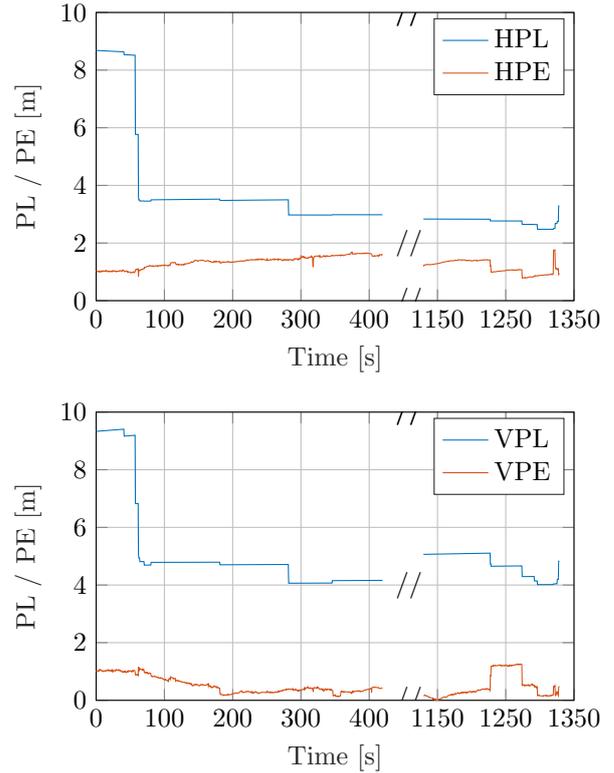}
    \caption{Real-time protection levels based on close-by GNSS reference during take-off and landing operations.}
    \label{fig:UGBAS_PE_PLs_live}
\end{figure}
\subsection{U-GBAS}\label{subsec:ugbas_test}
During test flights in July 2023 the U-GBAS subsystem was tested in real-time in terms of nominal performance for take-off and landing at close-by vertiport. The ground correction calculation and limited integrity monitoring was performed based on a single reference system consisting of a low-cost Tallysmann TW7972 triple band antenna, a Javad Omega receiver as well as a Laptop for processing and correction generation. The reference antenna was located about 100 m away from the vertiports. Nominal noise and multipath models for both ground and airborne antenna were derived from earlier tests using the same hardware \cite{gerb2023}. The system utilized L1/E1 measurements from up to 14 available GPS and Galileo satellites.

The flown trajectory where real-time computed U-GBAS positions and protection levels are available consists of two short flights between the two vertiports (see also \figref{fig:cochstedt}) at the Cochstedt test facility. \figref{fig:UGBAS_ref_trajectory} shows the trajectory generated from differential GNSS post-processing. This case, straight vertical take-off and landing were conducted, not considering future, potentially more complex approach trajectories. \figref{fig:UGBAS_PE_PLs_live} depicts the achieved protection levels as well as position errors (compared with the post-processing trajectory) for the two flights. 
As we can see, after an initial convergence phase where fewer satellites are used and therefore protection levels are higher, we reach values of 3-4~meters in horizontal and 4-5~meters in vertical domain, approximately in the order of magnitude of the requirements derived in section \ref{sec:nav_req}. 
While vertical results appear particularly promising in this favourable scenario (open sky, close-by reference), the achieved horizontal position errors show a bias of around 1~meter. 
\subsection{Ground-corrected barometer for vertiport operations}\label{subsec:baro_test}
\subsubsection{Setup and system calibration}
During the flight trials, the airborne pressure measurements were performed by the barometer of an industrial grade navigation system. A ground computer read the ground measurement from a weather station and transmitted them to the drone. The data was transmitted over the WLAN provided by a router and locally extended with an additional antenna. An airborne computer processed the airborne and ground pressure measurements to compute barometric geodetic altitude. 
\begin{figure}
    \centering
     \setlength\figwidth{0.87\linewidth}
     \setlength\figheight{1.2\linewidth}
    \input{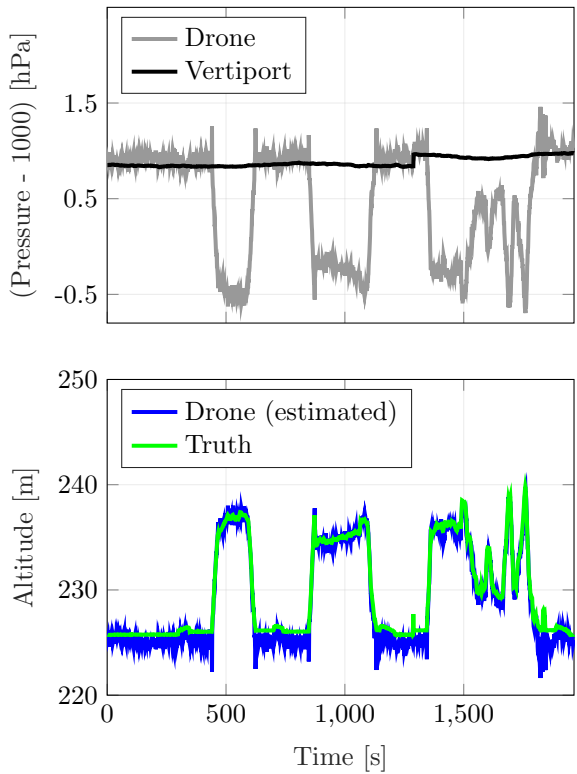}
    \caption{Pressure measurements (upper plot) and barometric geodetic altitude (lower plot) along HorizonUAM test flights.}
    \label{fig:baro_results}
\end{figure}
The employed airborne barometer was realistically of a notably lower grade if compared to barometers used in aviation or even to the one described in~\cite{Lerro2021} for UAM vehicles. For this reason an adjustment of the airborne barometer measurements was particularly needed. Biases could otherwise result in large altitude estimation errors. A comparison of a time series of pressure measurements of the adopted airborne barometer with the ones of the weather station showed an almost constant offset of approximately 193 Pa. A second comparison six weeks later showed an offset of 197 Pa. The employed airborne barometer and the weather station were factory\textendash calibrated several years ago and few months ago, respectively.
Based on this and on the specifications of these devices, it can be assumed the weather station to be more reliable. It can also be assumed that a constant bias of approximately 195 Pa, or a possibly slowly growing bias, is present in the airborne barometer. This bias would translate to an absolute altitude offset of approximately 16 m, at ISA MSL conditions. We hence adjusted the airborne pressure measurements by removing this bias, before employing them for the barometric geodetic altitude computation. 

In an operational scenario, we recommend the usage of reliable and stable ground weather stations. We also advocate regular inspections and calibrations of airborne barometers, in case their stability and accuracy will be limited by industry requirements for reduced cost and mass.
\subsubsection{Initial results}
The lower plot of \figref{fig:baro_results} presents some initial results of the barometric geodetic altitude computation based on the ground\textendash correction approach. The results represent several short flights over more than 30 minutes.
The upper plot of \figref{fig:baro_results} shows the airborne and the vertiport pressure measurements. From this plot, it can be noticed that the employed method produces altitude measurements that quite well capture the true altitude profile. However, the barometric geodetic altitude shows a quite noisy behavior. This is due to the noise in the airborne pressure measurements, of a notably higher level than the one in the ground pressure measurements. The highest errors in the geodetic altitude computation are found in those transients during which the drone is flying very close to the ground. Despite these errors, the proposed approach shows first promising performances as a vertical navigation aid for UAM applications.
\subsection{Camera-aided takeoff and landing}\label{subsec:camera_test}
\begin{figure}[t]
    \centering
    \includegraphics[width=.98\linewidth]{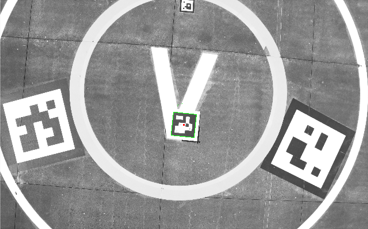}
    \caption{Marker detection (green line) during the landing phase at Vertiport 2.}
    \label{fig:marker_detection}
\end{figure}

It is very challenging to achieve the navigation requirements only using GNSS when vehicles fly close to the ground. Thus, vision-based positioning was applied to support the take-off and landing phases. 
For visual positioning, two different AprilTag markers \cite{olson2011tags,wang2016iros,krogius2019iros} were generated using the family type "tag25h9" and pattern ID=11 and 21. A marker with pattern ID=11 was placed on the take-off site (Vertiport 1 in \figref{fig:cochstedt}), and the other one with ID=21 was placed on the landing site (Vertiport 2 in \figref{fig:cochstedt}). Both markers' size was equally set as $0.785\mathrm{m}\times0.785\mathrm{m}$. \figref{fig:marker_detection} shows an example of the marker detection with the green line during the landing phase at Vertiport 2. 
%
%
\subsubsection{Error analysis of camera intrinsic parameter estimation} 
Before the flight experiments, the camera intrinsic parameters, such as focal length and the position of the optical center point, were estimated (camera calibration process). Then, the errors of the estimates were first analyzed since they have a significant impact on the camera position estimation. 
First, multiple ArUco markers \cite{aruco} were placed on the ground, and one of corner points was defined as the origin of the reference frame. True 3D locations of the other markers' corner points were computed with respect to the reference frame by measuring the distances between the markers. Then, corner points were detected in the sample images.
The 3D coordinates of each corner points were reconstructed with respect to the reference frame, using the camera intrinsic parameters (estimated during the calibration process) and the detected corner points. Finally, the differences between the reconstructed coordinates and the true 3D locations can be computed, and back-propagated to the intrinsic parameters. 
If the errors exceeds the required accuracy, the camera should be re-calibrated. Otherwise, the calibration results are accepted, and the uncertainty in the intrinsic parameters are propagated to the position estimates and included in the protection level calculation. The more detailed processes and results can be found in our paper \cite{hao2023camera}. 
%
%
\begin{figure}
    \centering
    \includegraphics[width=.98\linewidth]{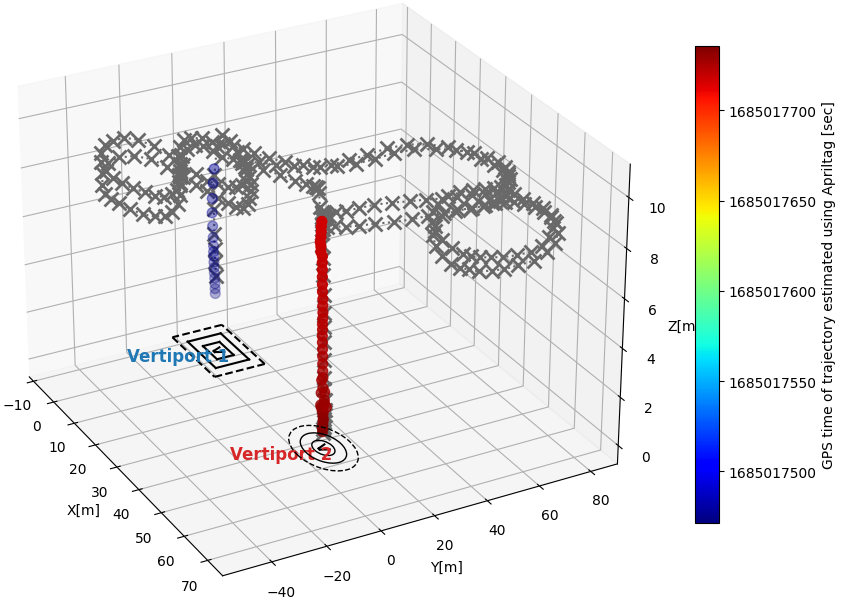}
    \caption{Vision-based trajectory estimates during take-off (Vertiport 1) and landing (Vertiport 2) phases (colorbar). Ground truth trajectory is depicted in gray.}
    \label{fig:trajectory_apriltag_vs_gnss}
\end{figure}
\subsubsection{Visual positioning validation}
In the experiments, the vehicle took off at the Vertiport 1, flew the planned trajectory, and then landed on the Vertiport 2. The ground truth trajectory (gray in \figref{fig:trajectory_apriltag_vs_gnss}) were obtained using the differential correction method with GNSS as well as GBAS signals, and the decimeter level accuracy is expected. 
During the take-off and landing phases, the 3D positions of the vehicle were estimated using the markers placed on each vertiport. The position estimates are shown with colorbar in \figref{fig:trajectory_apriltag_vs_gnss}. As can be seen in the figure, the vehicle's trajectory was accurately estimated using the marker-based method during the both phases, following the ground truth trajectory within meter level differences. 

%
\section{Outlook}\label{sec:outlook}
Future work will address the integration and further developments of the navigation subsystems and their validation and testing with respect to the foreseen navigation requirements for the intended operations.
The derivation of error and threat models for the UAM use cases will also be continued in future work. In particular, to reach a quantification of specific fault probabilities for each subsystem.
Additional sensors will also be considered in the future to support more reliable operations, for instance related to the final touchdown and taking off phase.

%
%
\section{Conclusions }\label{sec:conclusions}
This paper reviews current international standardization effort of UAM operations and derives first navigation requirements, with a focus on vertiport operations. The derived requirements will be instrumental for the design and validation of future UAM navigation systems and therefore for safe operations.

In this work, we present a brief overview of the available technical solutions for navigation systems suitable to UAM vertiport operations. We then focus on a subset of these solutions as subsystems of a first navigation integrity architecture proposal. Initial performance assessments of the chosen subsystems based on UAV flight trials show promising potential for their future inclusion within robust UAM navigation systems.
%
%
%
\backmatter
%
%
%
%
%
\section*{Acknowledgments}
This work has been performed in the context of the DLR internal project HorizonUAM. We would like to thank the aviation program directorate of DLR for their support. We would also like to thank Dennis Becker and Lukas Schalk for their support with the flight tests. 
\section*{Declarations}
%
\bmhead{Funding}
This work has been funded by the aviation program directorate of the German Aerospace Center (DLR) within the context of the HorizonUAM project.
\bmhead{Conflict of interest}
The authors declare neither conflict of interest or financial interest.
\bmhead{Ethics approval}
No human or animal has been involved in this research.
\bibliographystyle{IEEEtran}
\bibliography{bibliography/uam_paper, bibliography/uam}
\end{document}